# Research on topological materials using ultrafast spectroscopy*

LIU Hao, MENG Jianqiao

School of Physics, Central South University, Changsha 430083, China

**Abstract** Topological materials, characterized by symmetry-protected nontrivial band structures such as Dirac cones and Weyl nodes, exhibit a rich variety of quantum states and novel physical phenomena. These materials hold great promise for applications in quantum transport, spintronics, and nonlinear optics. In recent years, ultrafast pump-probe spectroscopy has become a powerful tool for studying nonequilibrium dynamics in quantum materials. With femtosecond temporal resolution, this technique enables direct observation of charge, spin, orbital, and lattice interactions on their intrinsic timescales, offering new insights into the coupling mechanisms in topological systems. This review summarizes the latest research progress of applying ultrafast spectroscopy to topological insulators, topological semimetals, and magnetic topological materials. We first discuss the different relaxation pathways of surface and bulk electronic states after photoexcitation, focusing on electron-phonon scattering, surface-bulk charge transfer, and ultrafast spin conversion. We then describe population inversion phenomena in Dirac and Weyl semimetals, spin polarization dynamics induced by tilted Weyl bands, and the influence of magnetic order on topological states, including coherent phonon and magnon excitations, magnetically driven topological transitions, and terahertz pulse generation. Furthermore, we review photoinduced topological phase transitions driven by electronic correlations, lattice distortions, and magnetic order under strong optical fields, highlighting the potential for nonthermal optical control of quantum phases. Finally, we discuss future research directions, emphasizing the integration of multidimensional ultrafast spectroscopic techniques—spanning temporal, energy, momentum, and spin resolution—with advanced theoretical simulations to construct a comprehensive picture of nonequilibrium topological states. This work aims to serve as a reference for studies on the ultrafast dynamics of topological quantum materials and to promote their practical applications in high-speed, low-power information processing, spintronics, and quantum computation.

---

* The paper is an English translated version of the original Chinese paper published in Acta Physica Sinica. Please cite the paper as: **LIU Hao, MENG Jianqiao, Research on topological materials using ultrafast spectroscopy.** *Acta Phys. Sin.*, 2026, 75(4): 040703. DOI: 10.7498/aps.75.20251330

# 1 Introduction

The electronic structure of topological materials are commonly characterized by symmetry-protected hourglass-like Dirac cones or Weyl cones. Their topological boundary states exhibit remarkable robustness against perturbations such as non-magnetic impurities, structural defects, and lattice deformations. This stability provides an ideal platform for realizing stable quantum transport and low-energy-consumption devices. With the rapid advancement of ultrafast laser technology, time resolution from the femtosecond (fs) to nanosecond (ns) regime now allow researchers to explore quasi-equilibrium and non-equilibrium dynamic processes in condensed matter systems, thereby advancing the time-domain characterization and dynamical manipulation of topological states of matter.

Novel phenomena such as polarization reversal[1], coupled lattice-magnetic-order excitations[2], and photoinduced phase transitions[3] have been discovered successively. These discoveries not only enrich the physical understanding of topological quantum states, but also open new routes toward programmable topological qubits and on-chip high-speed, low-power spintronic devices. Ultrafast spectroscopy reveals key processes such as electronic transitions, energy transfer, and multi-degree-of-freedom coupling by probing the relaxation dynamics of photo-excited carriers. Mechanisms such as electron-electron scattering and electron-phonon interactions often share similar characteristic energy scales. Therefore, effective resolution and decoupling in the time domain are essential for deeply understanding the intrinsic dynamic behaviors and topological response mechanisms of topological materials.

In recent years, ultrafast spectroscopy has achieved significant progress in systems such as topological insulators, topological semimetals, and magnetic topological materials[4-6]. Although these material systems differ in chemical composition and crystal structure, their shared topological band features often lead to similar ultrafast carrier dynamic behaviors. Given the vast variety of topological materials and the rapid, diverse development of ultrafast spectroscopy, it is impossible to review all research directions. This article therefore summarizes representative recent progress according to the following three principles. First, technical representativeness: we focus on the most commonly used laboratory-based techniques, namely reflection-type pump-probe ultrafast spectroscopy and time-resolved magneto-optical Kerr effect measurements. Second, material typicality: we select nonmagnetic and magnetic topological insulators and topological metals with clear evidence for topological properties. Third, clarity of physical mechanisms: we emphasize the coupled dynamics of electronic, lattice, and spin degrees of freedom induced by ultrafast optical pulses. The remainder of this review is organized as follows. We first introduce the basic principles and experimental methods of ultrafast spectroscopy. We then review its applications in representative topological material systems. Finally, we discuss the physical mechanisms of photoinduced topological phase transitions and provide an outlook on future opportunities and challenges.

# 2 Ultrafast Spectroscopy Techniques

Research on ultrafast carrier dynamics typically relies on pump-probe technology. This method uses a strong femtosecond laser pulse (pump beam) to excite the sample into a non-equilibrium state. Subsequently, under precisely adjustable time delay $t$, a weaker femtosecond laser pulse (probe beam) is introduced to monitor the process of the sample relaxing from the excited state back to equilibrium. The relaxation behavior depends on the intrinsic properties of the material and the pump photon energy. The relaxation process mainly involves interband and intraband transitions. During this process, various bosonic excitations, such as phonons, magnons, charge density waves, and spin density waves, may participate and play key roles[5-10].

Currently, the most widely applied technique is optical pump-optical probe (OPOP) spectroscopy. The wavelengths of its pump and probe pulses typically range from visible to near-infrared (approximately 400-1000 nm). By selecting different pump and probe wavelengths, researchers can obtain specific information about non-equilibrium states. Common experimental modes include degenerate OPOP, two-color OPOP, optical pump-continuum white-light probe spectroscopy, and ultrafast pump-probe spectroscopy extended to the terahertz spectral range.

A common measurement parameter in experiments is the transient change in sample reflectivity $\Delta R / R$ or transmissivity $\Delta T / T$ As shown in Fig. 1[11], ultrafast lasers are typically generated by femtosecond laser sources. Pump-probe spectroscopy often uses near-infrared pulsed light with a wavelength of approximately 800 nm. The pulse is split into pump and probe beams by a beam splitter, shaped, and then focused onto the sample surface. The optical path difference between the pump and probe beams is precisely adjusted via a delay stage, enabling signal acquisition with femtosecond time resolution. To extract effective signals from background noise, a chopper and a lock-in amplifier are usually used to modulate the pump beam. The AC photocurrent component of the reflected probe light is measured by a photodetector. This signal directly reflects the transient change in sample reflectivity under pump light excitation.

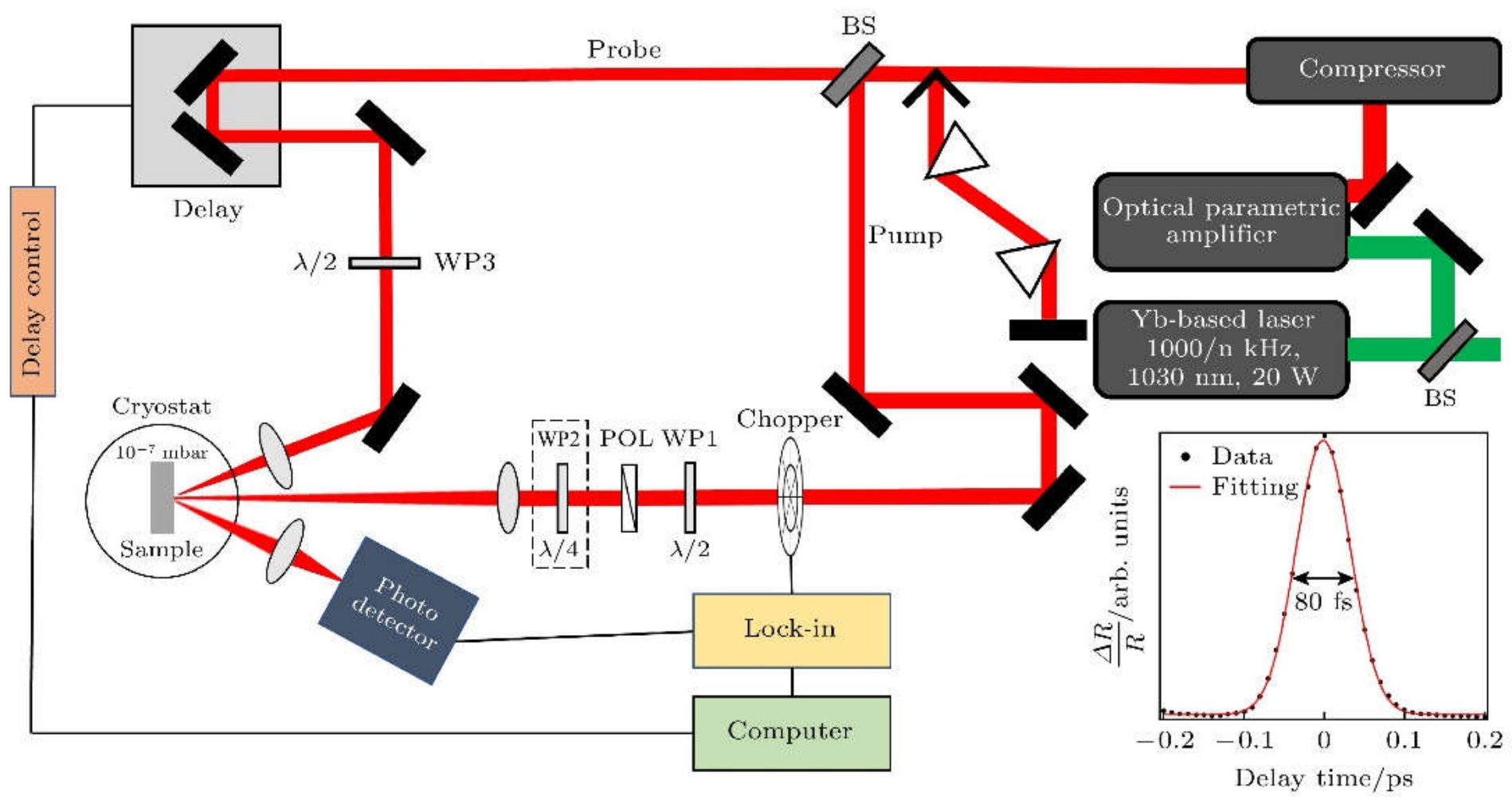

**Fig. 1 Schematic diagram of a typical ultrafast optical pump-optical probe setup[11].**

The changes in $\Delta R / R$ or $\Delta T / T$ originate from the pump-induced perturbation of the electronic distribution. As the system evolves from the excited state toward internal and external thermal equilibrium, changes in the electronic distribution modify the interband absorption spectrum and thus the optical response of the material[12]. The $\Delta R / R$ or $\Delta T / T$ signal of a sample can be expressed as a linear combination of changes in the real and imaginary parts of the complex dielectric function, $\Delta\epsilon_1$ and $\Delta\epsilon_2$. Using the Kramers-Kronig relations, it can be written in a form that depends only on the change in the imaginary part, $\Delta\epsilon_2$ [13]:

$$\frac{\Delta R}{R}=\frac{2}{\pi}\frac{\partial \ln R}{\partial \epsilon_1}PV\int_0^{\infty}d\omega'\frac{\omega'\Delta\epsilon_2(\omega')}{(\omega')^2-\omega^2}+\frac{\partial \ln R}{\partial \epsilon_2}\Delta\epsilon_2, \tag{1}$$

Here, PV denotes the Cauchy principal value integral.

Changes in the dielectric function are directly related to changes in the electronic distribution function and therefore need to be analyzed for each specific sample. For example, for gold, under the constant-matrix-element approximation, the change in the imaginary part of the dielectric function can be expressed in terms of the joint density of states and the change in the electronic distribution function [12,13]

$$\Delta\epsilon_2(\hbar\omega) = \frac{1}{(\hbar\omega)^2}\int D(E,\hbar\omega)\Delta\rho(E)dE,$$
$$D(E,\hbar\omega) = \frac{1}{(2\pi)^3}\int d^3k\,\delta[E_c(k) - E_v(k) - \hbar\omega]\times\delta[E - E_c(k)], \quad (2)$$

Here, $D(E,\hbar\omega)$ denotes the joint density of states at conduction-band energy $E$, obtained by calculating absorption transitions between the valence and conduction bands. Therefore, the temporal evolution of $\Delta R / R$ is determined, to some extent, by the band structure near the Fermi level. In ultrafast pump-probe experiments, femtosecond laser pulses have extremely short temporal widths and broad spectral bandwidths, allowing them to excite multiple elementary excitations simultaneously while maintaining well-defined phase relationships. This enables the selective excitation of coherent phonons and coherent magnons[14].

Coherent phonons refer to collective oscillations of lattice vibrational modes that maintain a fixed phase relationship in space and time. They can be classified into coherent optical phonons and coherent acoustic phonons. Coherent optical phonons are mainly generated through two mechanisms: impulsive stimulated Raman scattering (ISRS)[15] and displacive excitation of coherent phonons (DECP)[16]. The former originates from Raman-active modes in the crystal, whereas the latter arises from coherent atomic motion around a new equilibrium position after the crystal absorbs laser energy. These two mechanisms can be distinguished by the initial phase of the phonon oscillation[17]: under the commonly used convention for the lattice displacement, ISRS-driven oscillations are sine-like, whereas DECP-driven oscillations are cosine-like. In addition, coherent acoustic phonons mainly originate from thermoelastic stress waves induced by laser heating, deformation-potential coupling, or piezoelectric and magnetostrictive effects. They are widely observed in the low-frequency regime and can be used to probe interfacial phonon propagation and interlayer coupling dynamics [18-20].

After pump excitation, the material absorbs energy and forms a nonequilibrium population of photogenerated electrons. These high-energy electrons first undergo electron-electron thermalization, leading to a rapid redistribution of electronic occupation near the Fermi level. They subsequently transfer energy to other degrees of freedom through processes such as electron-phonon and electron-spin scattering. Within this framework, the system is usually divided into electronic, lattice, and spin subsystems, each assigned an effective temperature. This provides the basis for two-temperature or three-temperature models that describe energy-exchange dynamics[21-23].

Under high-fluence excitation, ultrafast laser pulses can also induce nonthermal metastable states with well-defined order parameters, or even trigger photoinduced quantum phase transitions, offering new experimental routes for studying nonequilibrium condensed matter systems.

With continuing technical advances, diverse experimental approaches have emerged. In addition to conventional $\Delta R / R$ and $\Delta T / T$ measurements, spin dynamics can be directly tracked by detecting pump-induced changes in optical polarization, such as Kerr rotation and ellipticity. This approach is known as ultrafast magneto-optical Kerr effect, or MOKE, spectroscopy. The spin polarization of a sample may originate from intrinsic ferromagnetism or magnetization induced by an external magnetic field[24,25]. It may also arise from chiral-charge nonconservation in special band structures excited by the pump pulse[1], or from spin injection induced by circularly polarized light[26,27].

Beyond magneto-optical detection, nonlinear optical processes provide a unique means of probing symmetry breaking. When an intense femtosecond laser pulse is incident on a noncentrosymmetric medium, the nonlinear polarization can generate a second-harmonic signal at twice the fundamental frequency. By symmetry, second-order nonlinear optical processes are forbidden in an ideal bulk crystal with inversion symmetry. Therefore, SHG signals mainly originate from surfaces, interfaces, or regions with local symmetry breaking. In recent years, time-resolved second-harmonic generation has become a highly sensitive and nondestructive probe of surface states and symmetry evolution in topological materials[28]. At higher energy and momentum resolution, time- and angle-resolved photoemission spectroscopy, Tr-ARPES, can directly map the relaxation of photoexcited electrons along electronic bands in momentum space[29]. Meanwhile, ultrafast terahertz spectroscopy, with photon energies on the meV scale, is particularly suitable for probing low-energy elementary excitations [8,30]. Ultrafast two-dimensional spectroscopy can reveal coherent coupling and relaxation between elementary excitations[31], whereas time-resolved resonant inelastic X-ray scattering (tr-RIXS) provides momentum- and energy-resolved information on collective excitations[32].

Overall, ultrafast spectroscopy covers multiple dimensions, from charge to spin, from linear to nonlinear optical responses, and from electronic bands to collective excitations. Different techniques have distinct advantages in time resolution, energy window, and detection sensitivity, and together they form a complementary experimental framework. Using these methods, researchers can reveal the dynamical coupling among electronic, lattice, and spin degrees of freedom in complex quantum materials, as well as the nature of their nonequilibrium evolution, on the intrinsic timescales of atomic and electronic motion.

# 3 Topological Insulators

## 3.1 Non-magnetic Topological Insulators

Topological insulators are a unique class of quantum matter in which the bulk is insulating, whereas the surfaces or edges host metallic, conducting electronic states protected by time-reversal symmetry ($\mathcal{T}$). These topological boundary states display Dirac-type band structures with linear dispersion and feature spin-momentum locking. This makes them promising for applications in spintronics and low-power devices. Among various topological material systems, the three-dimensional topological insulator $Bi_2(Se, Sb)_3$ family has become central platforms for ultrafast spectroscopic studies because of their relatively large bulk band gaps, well-defined surface states, and good materials stability. Ultrafast spectroscopy has shown that, under photoexcitation, carrier relaxation dynamics in topological insulators are strongly influenced not only by electron-phonon interactions but also by the presence of topological surface states. Ultrafast MOKE measurements have further captured spin dynamics coupled to Dirac surface states, providing direct evidence for understanding nonequilibrium evolution in systems with strong spin-orbit coupling.

Taking $Bi_2Se_3$ as an example, its band structure near the Fermi level is relatively simple. The bulk band gap is approximately 0.3 eV, and topologically protected Dirac surface states connect the conduction and valence bands within the bulk gap. As shown in Fig. 2(a1), room-temperature ultrafast spectroscopic measurements on $Bi_2Se_3$ single crystals show that photoexcited carriers undergo relaxation dominated by electron-optical-phonon and electron-acoustic-phonon scattering, accompanied by coherent phonon oscillations and thermal diffusion[33-37]. The coherent oscillatory signal arises from three optical phonon modes, $A_{1g}^1$, $E_g^2$, and $A_{1g}^2$, as shown in Fig. 2(a2), together with a low-frequency acoustic phonon mode. Among them, thermalized electrons mainly dissipate energy through scattering with the $A_{1g}^1$ phonon. Notably, the electron-phonon coupling strength extracted from ultrafast spectroscopy is usually larger than that obtained from spectroscopic measurements of the surface-state band. This discrepancy indicates that, in conventional pump-probe experiments, the detected signal mainly originates from the dynamical response of bulk carriers rather than purely from topological surface states[36-43]. Therefore, when interpreting ultrafast spectroscopic data, especially when discussing the distinctive dynamics of topological surface states, the bulk and surface contributions must be carefully disentangled.

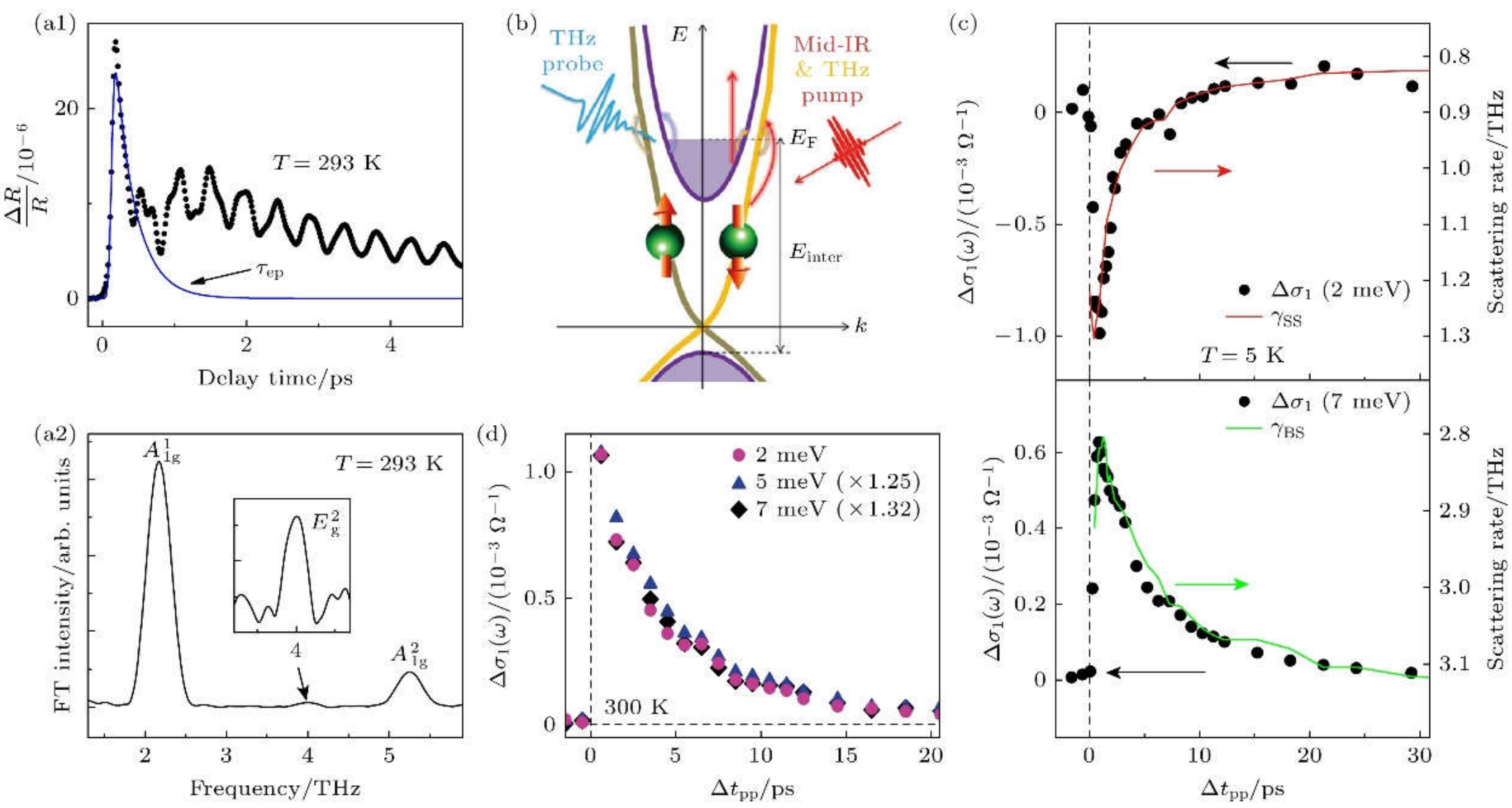

**Fig. 2 Ultrafast dynamics in $Bi_2Se_3$**: (a1), (a2) Transient reflectivity $\Delta R / R$ measured at $h\nu = 1.55$ eV and the fast Fourier transform spectra of the coherent oscillations[26]; (b) schematic of the surface and bulk band structures and the mid-infrared/terahertz pump-terahertz probe scheme[44]; (c), (d) terahertz conductivity as a function of time at 5 K and 300 K, respectively, measured with probe photon energies of 2, 5, and 7 meV. At low5 temperature, surface-state quasiparticles relax faster than bulk-state quasiparticles; at high temperature, enhanced coupling leads to nearly identical relaxation times[44].

In nonequilibrium dynamics, the selective excitation of specific band transitions in topological insulators using mid-infrared or terahertz pulses, combined with terahertz probing of low-energy excitations, can effectively distinguish the dynamical responses of bulk and surface states, as illustrated in Fig. 2(b)[44-48]. This approach benefits from the high sensitivity of terahertz radiation to carrier conductivity and is particularly suitable for probing gapless Dirac fermions in topological surface states. In addition, when $Bi_2Se_3$ is thinned from bulk material to thin films, the relative contribution from metallic surface states is substantially enhanced, leading to faster overall carrier relaxation. Therefore, ultrafast spectroscopy on thin-film samples is useful for isolating and identifying the role of Dirac surface states in photoinduced electronic responses [49,50]. Mid-infrared ultrafast spectroscopy shows that bulk states mainly undergo intraband thermalization governed by electron-optical-phonon scattering, together with long-lived electron-hole recombination across the bulk band gap[44,51]. By contrast, ultrafast terahertz spectroscopy reveals that, at low temperatures, the relaxation time of the surface-state conductivity change associated only with intraband transitions is significantly shorter than that of the bulk states, as shown in Fig. 2(c). At high temperature, however, phonon-assisted bulk-surface coupling channels become active, allowing a large number of bulk electrons to relax rapidly through the surface states. As a result, the relaxation behaviors of bulk and surface states become similar, as shown in Fig. 2(d)[44,45,52].

Regarding spin dynamics, ultrafast magneto-optical Kerr effect measurements and surface-sensitive second-harmonic generation spectroscopy provide powerful tools for tracking the evolution of spin polarization in $Bi_2Se_3$. Owing to the helical spin texture of topological surface states, bulk and surface states exhibit distinct spin relaxation behaviors after spin-polarized carriers are injected by circularly polarized pump light [26,27,53,54], as shown in Figs 3(a) and 3(b). Surface-state electrons, with relaxation time denoted by $\tau_s$, are mainly affected by $E_g^2$ phonon scattering and exhibit a faster spin relaxation rate than bulk electrons, whose relaxation time is denoted by $\tau_b$ and whose dynamics are mainly governed by $A_{1g}$ phonon scattering, as indicated in Figs 2(a1) and 2(a2)[26]. This difference is attributed to stronger spin-orbit coupling in the surface states and to differences in the momentum-dependent scattering phase space [26,27].

In summary, the charge and spin dynamics of surface states after photoexcitation can be described by the processes shown in Fig. 3(c1)-(c4)[27,55]. Circularly polarized pump light first excites spin-polarized carrier populations in both the bulk and surface states. Under strong spin-orbit coupling, these carriers rapidly undergo spin depolarization. Subsequently, within approximately 1 ps, surface-state electrons complete intraband cooling through electron-electron and electron-phonon scattering. Finally, the system enters a slower relaxation stage dominated by electron-hole recombination and thermal diffusion mediated by electron-acoustic-phonon scattering. It is worth noting that when high-energy photons excite electrons into the unoccupied second surface state above the Fermi level, their recombination with holes near the Fermi level must satisfy spin-conservation constraints. Consequently, the relaxation time of this process is significantly longer than that of simple electron-phonon-coupling-induced cooling, demonstrating the role of spin degrees of freedom in controlling the lifetime of nonequilibrium carriers[49,50].

Similar behaviors have also been observed by ultrafast spectroscopy in other $Bi_2Se_3$ family topological insulators, including $Bi_2Te_3$[56-59], $Bi_{1.5}Sb_{0.5}Te_{1.7}Se_{1.3}$[40,60-62], $Bi_2Te_2Se$[47,63], and $Sb_2Te_3$[64-66]. These studies show charge and spin relaxation processes dominated by electron-phonon coupling, accompanied by pronounced coherent phonon oscillations.

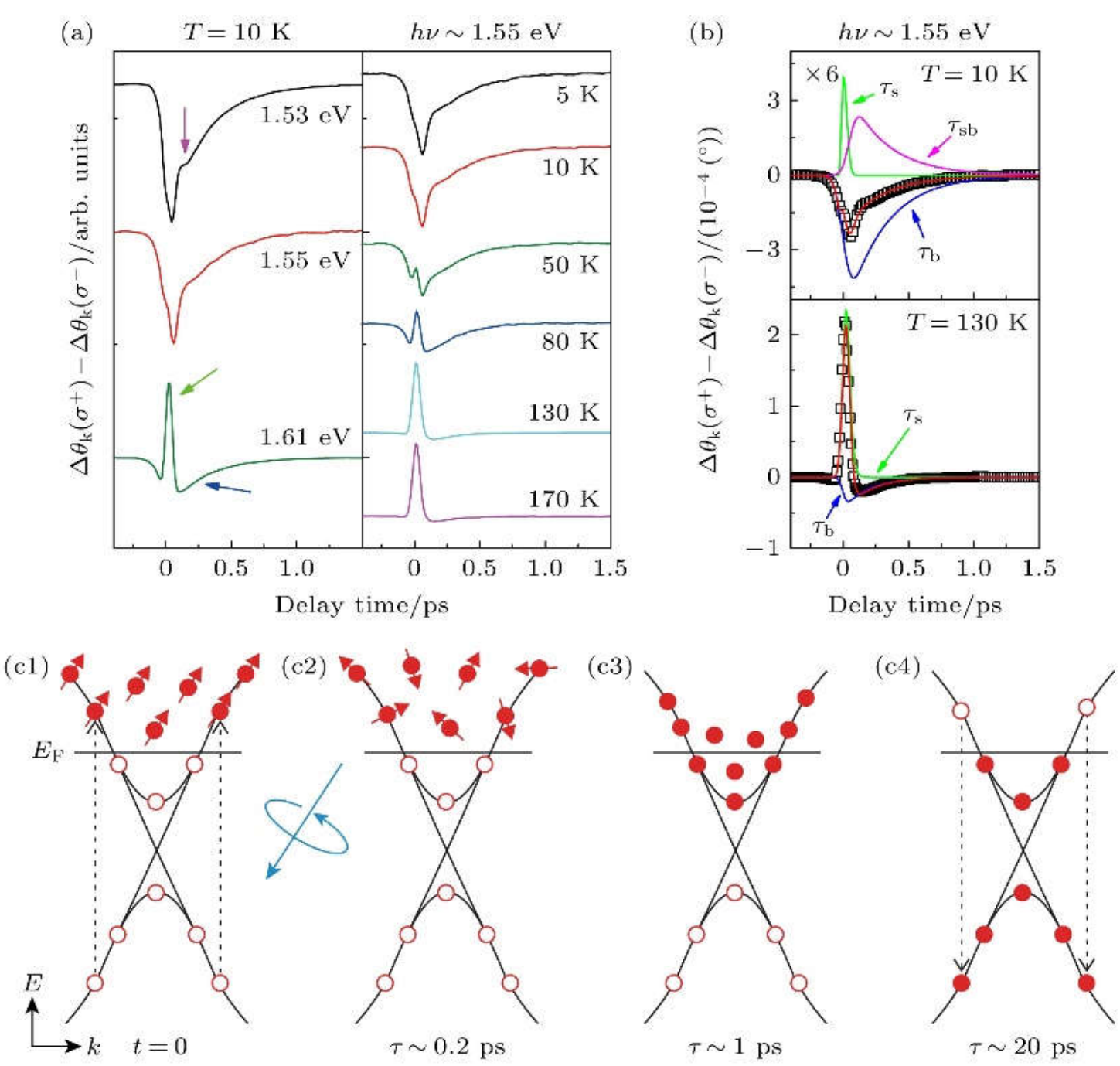

**Fig. 3 Spin dynamics in $Bi_2Se_3$:** (a) Time evolution of the magneto-optical Kerr angle at different photon energies and temperatures[26]; (b) triple-exponential fitting results corresponding to relaxation in the surface states ( $\tau_s$ ), relaxation in the bulk states ( $\tau_b$ ), and transfer from the surface states to the bulk states ( $\tau_{sb}$ )[26]; (c) schematic illustration of the ultrafast relaxation of surface charge and spin under circularly polarized excitation [27].

## 3.2 Magnetic Topological Insulators

The introduction of magnetic atoms breaks time-reversal symmetry and thereby opens a magnetic exchange gap at the Dirac point of the surface states. Van der Waals heterostructures formed by combining topological insulators with two-dimensional materials, such as transition-metal dichalcogenides and magnetic materials, not only provide a platform for exploring the interplay among topological order, charge order, and magnetic order, but also open new routes for manipulating Dirac surface states[67-70]. For example, in 2013, the quantum anomalous Hall effect was first realized in Cr-doped (Bi, Sb)$_2$Te$_3$ thin films at 30 mK[71]. The perfect quantization and vanishing longitudinal resistance achieved under such extreme conditions have stimulated extensive discussion on the coupling mechanism and coupling strength between magnetism and surface states.

Ultrafast spectroscopic studies of topological-insulator heterostructure thin films have revealed efficient spin-to-charge conversion at heterointerfaces[33,69,72-75].

Specifically, ultrafast magneto-optical Kerr effect measurements have shown that orbital hybridization between topological surface states and ferromagnetic layers can substantially accelerate ultrafast demagnetization[76-78]. Terahertz emission spectroscopy further confirmed that, at the $SnBi_2Te_4$/Co heterojunction, ultrafast spin injection and spin-to-charge conversion occur primarily at the surface and are dominated by topological surface states[74]. Liu *et al.*[79] investigated the ultrafast magnetization dynamics of Cr-$(Bi, Sb)_2Te_3$/CrSb heterojunctions using ultrafast MOKE. They found that the interface in contact with the antiferromagnetic CrSb layer can generate a photoinduced transient magnetic phase transition with an antiparallel spin configuration, mediated by exchange coupling through itinerant Dirac fermions. These results demonstrate strong coupling between surface states and substrate magnetism in magnetic topological heterostructures. Notably, ultrafast THz pulse emission has been realized in antiferromagnet-topological-insulator heterostructures such as MnSe/$(Bi, Se)_2Te_3$. Optimized circularly polarized THz emission can be achieved through the inverse spin Hall effect and laser-induced transient magnetic moments[75]. Moreover, active control of the THz polarization state by tuning the pump polarization provides a new route toward programmable THz light sources[75]. This is not only a successful application of topological materials in optoelectronic devices, but also an advance that promotes the development of ultrafast THz spectroscopy.

$MnBi_{2n}Te_{3n+1}$ represents a family of layered topological insulators with intrinsic magnetism. Its crystal structure consists of alternately stacked $MnBi_2Te_4$ and $Bi_2Te_3$ units, providing a rich physical landscape in which magnetic order coexists with topological states. In thin-film systems of this family, novel quantum phenomena such as the quantum anomalous Hall effect, axion-insulator states, and magnetic-field-driven topological phase transitions have been realized[80-85]. For instance, by designing $MnBi_2Te_4$-based heterostructures, the antiferromagnetic quantum anomalous Hall effect can be precisely tuned in a multi-parameter space, revealing cascaded quantum phase transitions driven by spin-configuration changes and a significant enhancement of quantum transport by in-plane magnetic fields[84]. The tunable coupling between spin and other microscopic degrees of freedom, such as charge, lattice, and orbital degrees of freedom, is regarded as a key physical basis for topological quantum control and multifunctional device integration. Although many ARPES studies suggest that the relationship between the surface-state gap and magnetic order in bulk $MnBi_2Te_4$ remains controversial[86,87], ultrafast spectroscopy provides an independent and complementary perspective for revealing many-body interactions in its nonequilibrium dynamics on picosecond-to-nanosecond timescales.

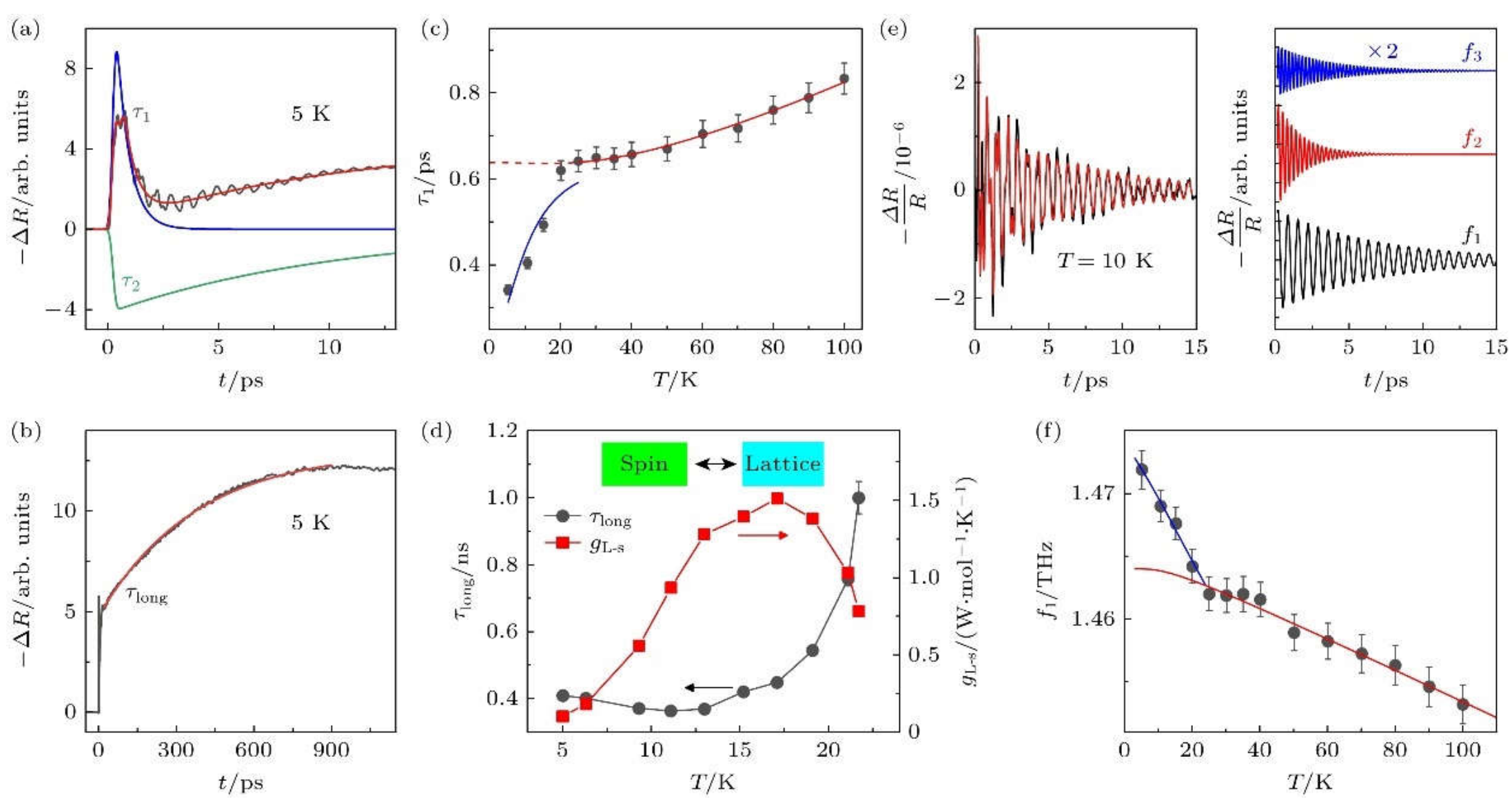

**Fig. 4 Ultrafast dynamics of $MnBi_2Te_4$ crystal**[88]: (a), (b) Transient reflectivity $\Delta R / R$ and fitting curves; (c), (d) temperature dependence of the relaxation time $\tau_1$ and $\tau_{long}$ coherent phonon frequency; (e) oscillatory component extracted at 10 K, , shown in black, together with the red fitting curve obtained using damped harmonic oscillators, and the three fitted oscillatory components at different frequencies; (f) temperature dependence of $A_{1g}^1$ coherent phonon frequency $f_1$.

In $MnBi_2Te_4$ single crystals, the relaxation behavior of non-equilibrium carriers and coherent phonon responses are significantly influenced by antiferromagnetic order ($T_N$ ~ 25 K). Moreover, a long-duration rise process appears after approximately 10 ps, as shown in Fig. 4(a) and Fig. 4(b). When the temperature is below $T_N$, the relaxation time $\tau_1(T)$ deviates from predictions of the classical two-temperature model. The relaxation time $\tau_{long}$ for energy exchange between spin and lattice subsystems, also exhibits divergent behavior near $T_N$, as shown in Figs. 4(c) and Fig. 4(d). Meanwhile, the frequency $f_1$ of the $A_{1g}^1$ coherent phonon shows anomalous hardening below $T_N$, as shown in Figs. 4(e) and 4(f)[88]. Under an external magnetic field, the variation of the $A_{1g}$ phonon amplitude is highly consistent with the suppression of antiferromagnetic order[89]. These experimental results indicate that coherent phonons can sensitively reflect the strong coupling between degrees of freedom such as charge and lattice and the antiferromagnetic order[88-90]. Furthermore, ultrafast magneto-optical Kerr effect and resonant soft X-ray scattering experiments have identified distinct ultrafast demagnetization processes on the picosecond time scale of electron-optical phonon interactions. Padmanabhan *et al*.[91] proposed that, under strong exchange coupling, the disordering of itinerant *p*-type spins can in turn induce simultaneous demagnetization of localized spin-polarized 3*d*-orbital states. Ultrafast spectroscopy has therefore established, on the femtosecond timescale, strong exchange-coupled interactions among charge, lattice, and antiferromagnetic order in $MnBi_2Te_4$ single crystals[88-92].

Theoretical calculations further indicate that, in bilayer $MnBi_2Te_4$ or $MnSb_2Te_4$ thin films, optical excitation can modulate the average displacement of the $A_{1g}$ Raman "breathing" mode, thereby changing the sign of the interlayer exchange interaction. This can induce a transition from an antiferromagnetic state to a ferromagnetic state and trigger a topological phase transition[92]. Experimentally, ultrafast spectroscopy has shown that carrier relaxation and coherent acoustic phonon responses in $MnBi_2Te_4$ thin films are highly sensitive to sample thickness[2,93,94]. Owing to the symmetry difference between antiferromagnetic sublattices in few-layer films, even-layer films show pronounced changes in optical response under relatively low external magnetic fields. Particularly noteworthy is that ultrafast spectroscopy recently captured dynamical axion quasiparticles for the first time in the two-dimensional antiferromagnetic topological material $MnBi_2Te_4$. The corresponding axion angle $\theta$, which is proportional to the magnetoelectric coupling coefficient, exhibits periodic oscillations at approximately 44 GHz. This behavior is attributed to ultrafast coherent modulation of the real-space Berry-curvature dipole induced by magnons[85].

Furthermore, when a strong in-plane magnetic field is applied, the even-layer $MnBi_2Te_4$ system enters a canted antiferromagnetic phase. Ultrafast spectroscopy successfully detects coherent magnon signals whose frequencies vary continuously with magnetic field, in excellent agreement with theoretical calculations[2]. Under an out-of-plane magnetic field, the ultrafast MOKE signal displays four characteristic features. First, the initial step-like response corresponds to transient demagnetization caused by electron-lattice thermalization. Second, driven by lattice-spin interactions, the magnetic susceptibility slowly decreases to the appropriate thermal-equilibrium value on an approximately nanosecond timescale. Third, the thermally equilibrated system cools back to ambient temperature over several nanoseconds through thermal diffusion and multistage phonon-phonon scattering. Finally, the accumulation effect of pump pulses produces a long-lived background offset that can persist for up to 100 ns[2]. These results indicate that magnon-phonon coupling is not only central to understanding the intrinsic dynamics of magnetic topological materials, but also provides a route for ultrafast manipulation of magnetic topological phases.

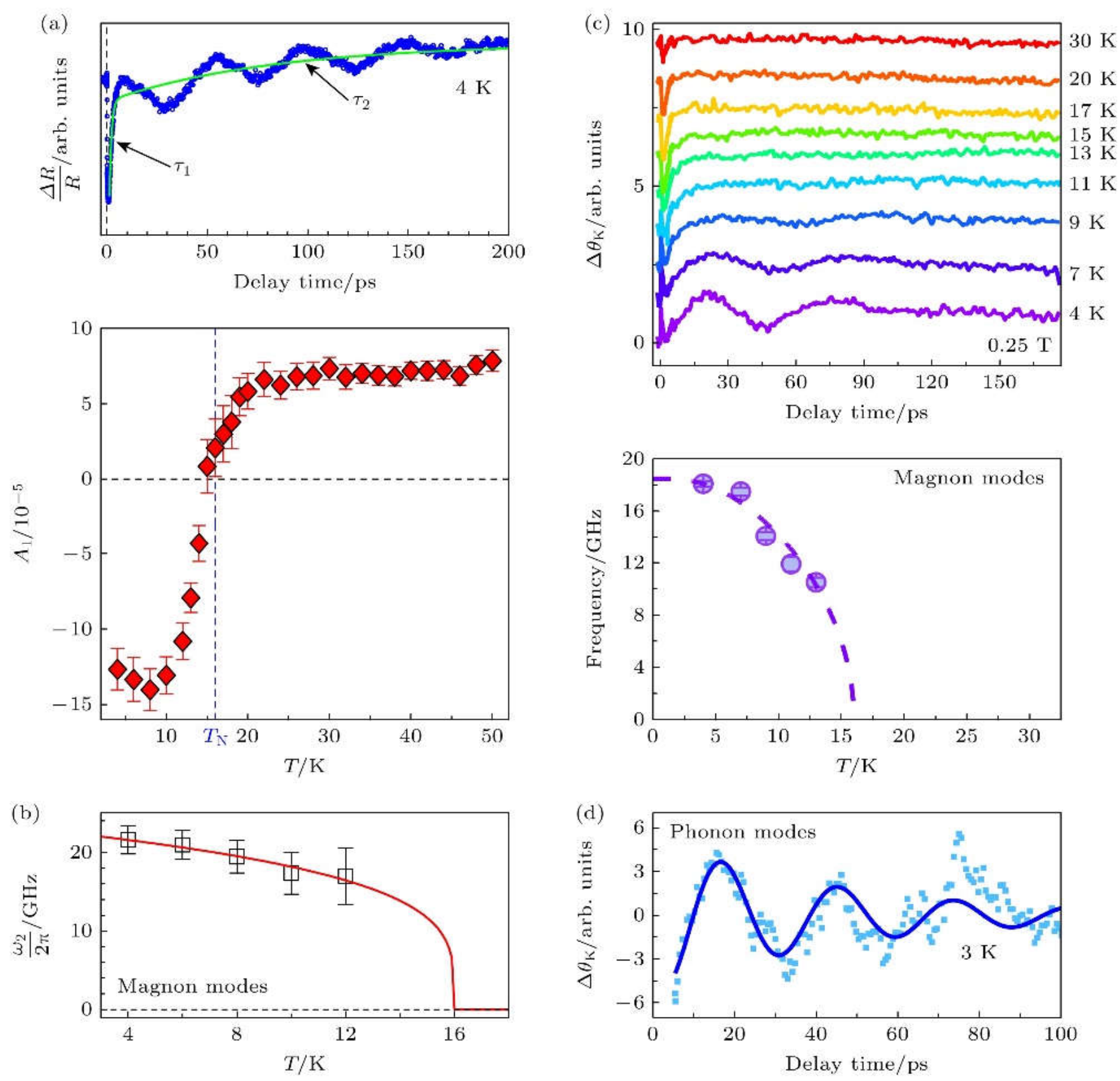

**Fig. 5 Ultrafast dynamics in $EuIn_2As_2$**: (a) Transient reflectivity $\Delta R / R$ at low temperatures and the temperature dependence of amplitude $A_1$, showing a sign reversal near $T_N$[97]; (b) temperature-dependent magnon modes extracted from reflectivity measurements[97]; (c) magneto-optical Kerr angle measured at different temperatures under a low magnetic field, together with the temperature dependence of the magnon frequency[98]; (d) phonon modes detected by MOKE measurements at 2 T[98].

Beyond the $MnBi_{2n}Te_{3n+1}$ family, Eu-based topological materials with strong spin-orbit coupling and antiferromagnetic order, such as $EuIn_2As_2$, have also become ideal platforms for studying the coupling between magnetism and topological states. Theoretical studies predict $EuIn_2As_2$ to be an axion insulator with topological invariant $Z_4 = 2$[95]. ARPES experiments have revealed bulk band inversion near the Fermi level, together with pronounced band reconstruction upon entering the antiferromagnetic phase[96]. Liu *et al.*[97] used ultrafast spectroscopy to reveal strong correlations between magnetic order and topological bands in this material. As shown in Fig. 5(a), the relaxation amplitude of the transient reflectivity signal reverses sign near the antiferromagnetic transition temperature $T_N$, indicating that the ultrafast dynamics of topological charge are strongly influenced by magnetic order. Both ultrafast degenerate pump-probe spectroscopy, Fig. 5(b), and MOKE measurements, Fig. 5(c), observed temperature-dependent coherent magnon excitations below $T_N$. These excitations originate from the collective spin precession of $Eu^{2+}$ magnetic moments[97,98]. In addition, Liu *et al*.[97] observed, in degenerate OPOP measurements, an extremely low-frequency coherent oscillation that cannot be explained as a coherent acoustic phonon. Because this oscillation appears only in the antiferromagnetic phase, it implies coupling between spin and lattice degrees of freedom at low temperature. Wu *et al.*[98] also detected a coherent phonon mode with a frequency of approximately 35 GHz in ultrafast MOKE experiments under strong external magnetic fields above 1.4 T, as shown in Fig. 5(d). Its generation mechanism is associated with a transient perturbation of the magneto-optical coefficient. These findings reveal the complex interplay among topological charge, spin, and lattice degrees of freedom in $EuIn_2As_2$.

# 4 Topological Semimetals

Topological semimetals are a class of quantum materials with topologically protected band structures. Their conduction and valence bands cross near the Fermi level, forming gapless degenerate points or lines. These degeneracies are protected by time-reversal symmetry ($\mathcal{T}$), spatial inversion symmetry ($\mathcal{P}$), mirror symmetry, or nonsymmorphic symmetries, and therefore exhibit high stability[99-101]. According to low-energy effective models, topological semimetals can be classified into Dirac semimetals, Weyl semimetals, and nodal-line semimetals. Owing to the strongly enhanced Berry curvature near Dirac or Weyl nodes, topological semimetals exhibit a range of novel electrical and thermal transport phenomena, including the anomalous Hall effect, the anomalous Nernst effect, negative magnetoresistance, and nonsaturating large magnetoresistance associated with the chiral anomaly[102,103]. These properties make topological semimetals an important platform for exploring exotic quantum effects and for developing low-power electronic devices.

The relaxation processes of photoexcited carriers in topological semimetals differ from those in topological insulators. Although they exhibit metallic behavior in certain aspects, their unique band structures give rise to strong nonlinear optical responses [104,105]. Photoexcited electrons undergo multiple ultrafast relaxation processes, including electron-electron scattering, electron-phonon interactions, electron-hole recombination, and phonon-phonon scattering. These processes relax carriers to the vicinity of the Fermi surface, forming hot electrons that satisfy the Fermi-Dirac distribution[106]. Because of the low density of states near the Fermi level, the electron-hole recombination rate is strongly reduced by phase-space constraints[105]. If the electron-hole recombination time exceeds the thermalization time of photoexcited carriers, electrons and holes accumulate near the band nodes. This accumulation forms Fermi-Dirac distributions with different chemical potentials[107]. The resulting population inversion may support broadband lasing beyond the mid-infrared region. In particular, when a gap opens at the nodes, the long-lived population inversion produced by phase-space blocking and phonon-bottleneck effects provides an ideal platform for studying excitonic-insulator states and high-harmonic generation[105,107-110].

At present, most ultrafast pump-probe experiments utilize lasers with a wavelength of 800 nm (equivalent to 1.55 eV). This photon energy can also excite electrons in non-Dirac bands. Although some photoexcited carriers relax through Dirac bands, experimental signals often contain contributions from other relaxation channels. To study the relaxation dynamics of Dirac and Weyl fermions more precisely, an improved strategy is to use mid-infrared or terahertz pulses to selectively excite carriers in the Dirac or Weyl bands[44-48]. This approach not only avoids interference from nontarget bands, but also reveals ultrafast dynamical processes that are intrinsic to topological semimetals.

## 4.1 Dirac Semimetals

Dirac semimetals possess many unique physical properties, including linear band dispersion, pronounced Berry curvature, an almost vanishing density of states near the Fermi level, and sensitivity to crystal symmetry and time-reversal symmetry. These characteristics are closely related to their optical responses on subpicosecond timescales[104,107]. Ultrafast laser pulses can break time-reversal symmetry or spatial inversion symmetry, thereby modifying the Dirac semimetal state[111].

In graphene films, the population inversion persists for approximately 100-200 fs, followed by an electron-hole recombination process lasting several picoseconds[108,112]. A similar behavior is observed in the three-dimensional Dirac material $Cd_3As_2$[113-119]. After photoexcitation, carrier thermalization first occurs within about 400-500 fs and is mainly driven by intraband relaxation dominated by electron-phonon scattering. This is followed by electron-hole recombination on a timescale of several picoseconds, accompanied by optical phonon emission. Because phase-space scattering is restricted, photoexcited carriers accumulate in the conduction band with a lifetime of about 3 ps, thereby forming a population inversion[117]. Ultrafast spectroscopy experiments on $Cd_3As_2$ films using mid-infrared photons, 2 μm corresponding to 0.62 eV, as the probe have revealed more specific dynamics[114]. Under high-energy 800 nm pumping, the initial response remains nearly flat for the first few hundred femtoseconds, indicating thermalization of non-Dirac electrons. By contrast, degenerate mid-infrared measurements directly capture the rapid thermalization and electron-hole recombination of Dirac fermions. After the plateau response, the relaxation curves overlap, showing that probe wavelengths close to the Dirac-cone energy can effectively track the dynamics of Dirac fermions[114].

Because bulk and nanosheet samples have different doping levels, the sign and amplitude of the transient reflectivity signal $\Delta R / R$ differ markedly. For example, bulk samples are highly doped, with the Fermi level approximately 200 meV above the Dirac point, whereas nanosheets are weakly doped, with the Fermi level about 38 meV above the Dirac point[118]. Consequently, the photoexcited carrier transitions and relaxation processes involved in bulk samples also involve non-Dirac bands. As shown in Fig. 6(a), under low pump fluence, $Cd_3As_2$ exhibits a negative $\Delta R / R$ response, which is mainly caused by changes in transitions between Dirac bands, indicated by the brown arrows in Fig. 6(b)[120]. Under high pump fluence, $\Delta R / R$ shows an enhanced positive response at the initial stage because of transitions involving non-Dirac bands, indicated by the green arrows in Fig. 6(b). As high-energy carriers relax back to the Dirac bands, $\Delta R / R$ changes into a negative response[120].

In Dirac semimetals, coherent phonons not only participate in electron-hole recombination under population-inverted conditions, but are also closely associated with ultrafast-laser-induced topological phase transitions[121]. Ultrafast pump-probe spectroscopy at 800 nm on $Cd_3As_2$ single crystals revealed beating among three coherent optical phonon modes induced by helical Cd-vacancy order, as shown in Fig. 6(c). With increasing temperature, the amplitudes of the phonon modes associated with the helical vacancies decay rapidly, except for the central $A_{1g}$ phonon mode. This suggests that temperature-driven lattice evolution may influence the topological properties of $Cd_3As_2$ (Fig. 6(d)).

Dirac semimetals possess ultrahigh mobility and a zero bandgap. In $Cd_3As_2$ thin films, laser-induced non-equilibrium carrier temperature gradients can generate picosecond thermoelectric currents and terahertz radiation[119,122]. In addition, under terahertz-field driving, the Dirac-electron distribution in $Cd_3As_2$ can be strongly stretched and asymmetrically distorted, leading to high-harmonic generation[123]. Zhu *et al*.[124] used time-resolved optical-pump terahertz-harmonic-generation spectroscopy to track terahertz third-harmonic generation in $Cd_3As_2$ thin films. They found that the long-lived population inversion generated by photoexcitation makes a substantial contribution to this phenomenon, as shown in Fig. 6(e). Theoretical analysis further indicates that the efficiency of terahertz third-harmonic generation increases significantly when the lifetime of the population-inverted state exceeds the terahertz pulse period, approximately 2.9 ps[124].

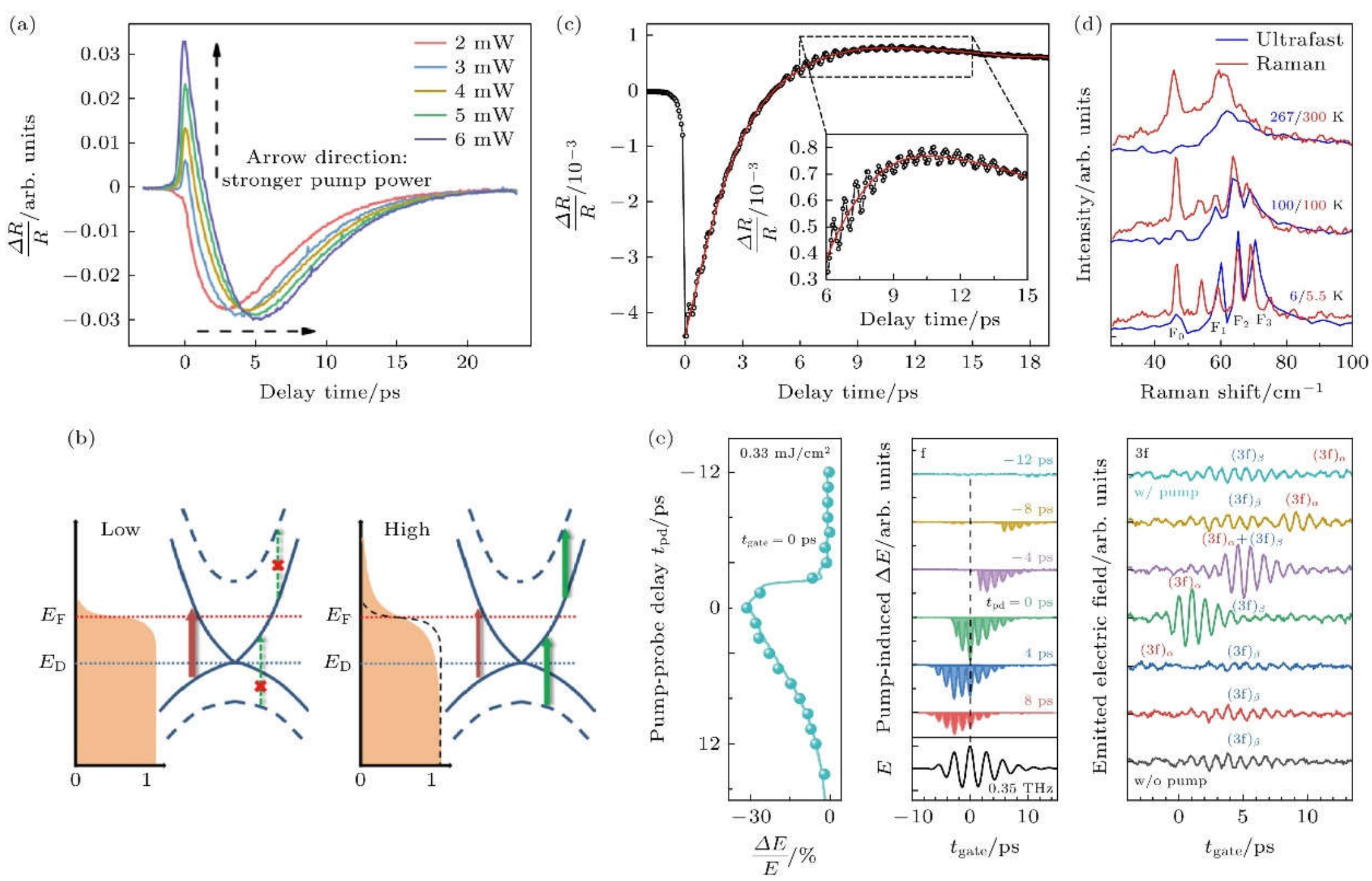

**Fig. 6 Ultrafast dynamics in the topological semimetal $Cd_3As_2$**: (a) Transient reflectivity measured with an 800 nm pump and a 4 μm probe at different fluences, showing an enhanced positive response at high fluence[120]; (b) schematic illustration of thermalized-electron redistribution and possible transition channels at low (left) and high (right) fluence[120]; (c) transient reflectivity measured at 6 K using 800 nm pump-800 nm probe spectroscopy, showing pronounced oscillations[121]; (d) coherent phonon modes at different temperatures obtained from fast Fourier transform and Raman spectra[121]; (e) relationship between the intensity of terahertz third-harmonic generation and the photoinduced population inversion[124].

Type-II Dirac semimetals possess tilted Dirac cones, leading to the coexistence of electron and hole pockets near the Fermi level. This electronic structure facilitates scattering between photoexcited carriers and the lattice. As shown in Figs. 7(a) and 7(b), ultrafast pump-probe spectroscopy has revealed an $A_{1g}$ phonon-assisted electron-hole recombination process in $NiTe_2$[125]. The relaxation time $\tau_2$ of this process, together with the frequency and decay rate of the $A_{1g}$ phonon, shows nonmonotonic behavior near the Lifshitz transition temperature $T^*$. These results reveal an ultrafast carrier-phonon coupling mechanism regulated by the Lifshitz transition.

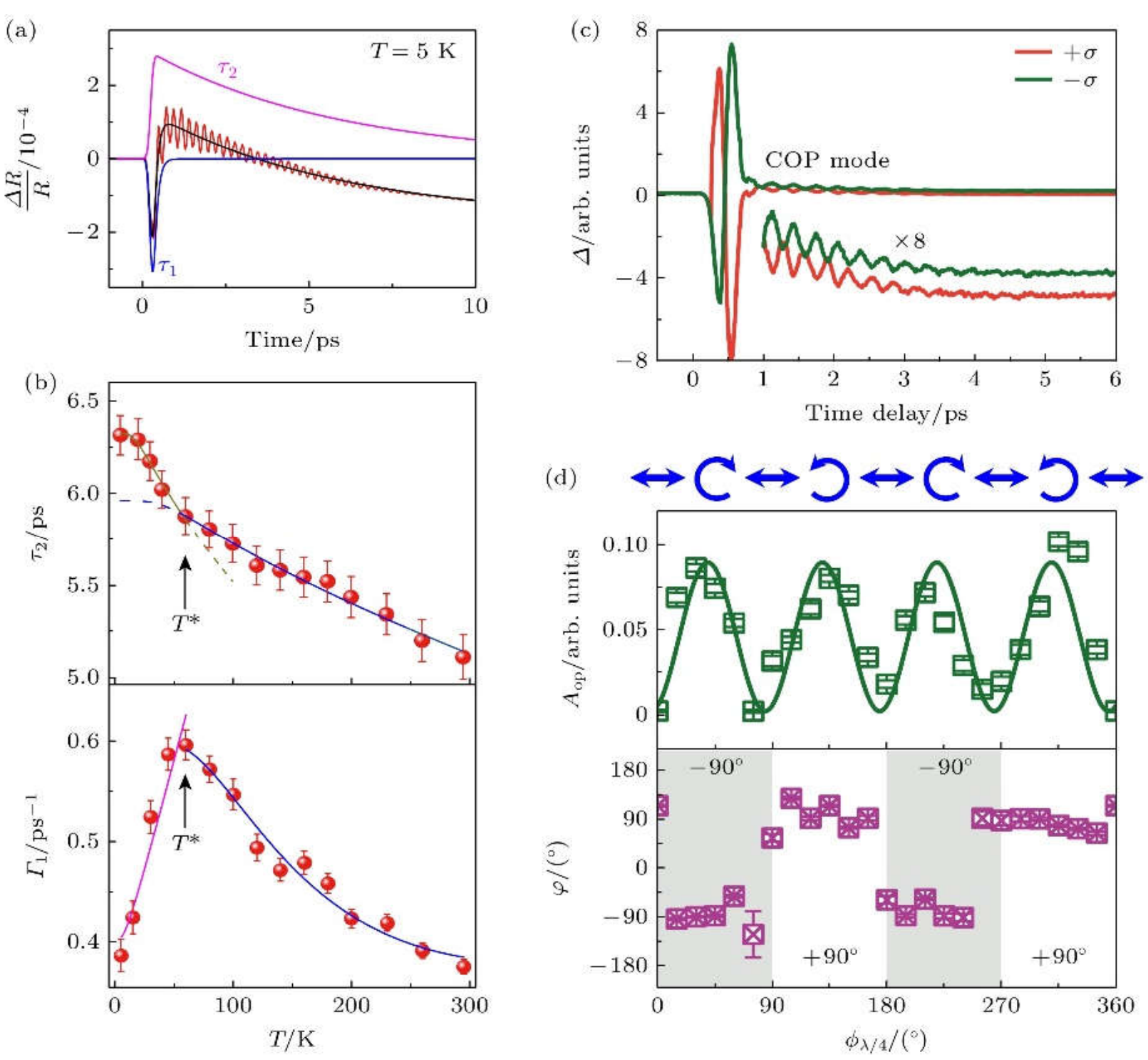

**Fig. 7 Ultrafast dynamics of Type-II Dirac semimetals**: (a), (b) Transient reflectivity of $NiTe_2$ under 800 nm excitation and the temperature dependence of the electron-hole recombination time $\tau_2$ and the coherent $A_{1g}$ phonon decay rate, showing an abrupt change at $T^* = 60$ K[125]; (c), (d) ultrafast differential signals in $PtTe_2$ thin films under different circular polarizations, where the amplitude and phase of the coherent oscillations vary with the helicity of the pump light[128].

In the type-II Dirac semimetal $PtTe_2$, coherent phonon signals corresponding to the in-plane $E_g$ mode and the out-of-plane $A_{1g}$ mode have been observed, in addition to electron-hole recombination processes[126,127]. Notably, in $PtTe_2$ thin films, angular momentum transfer from photoinjected spin-polarized electrons to the lattice induces asymmetric displacement of Te atoms in the in-plane $E_g$ mode. As a result, the amplitude and phase of this mode depend on the circular polarization of the pump light, as shown in Figs. 7(c) and 7(d)[128]. Optical-pump terahertz-probe experiments on $PtTe_2$ thin films have revealed an anomalous negative terahertz photoconductivity, which is attributed to the formation of small polarons induced by strong electron-phonon coupling[129]. Specifically, after photoexcitation, strong electron-optical-phonon coupling produces local lattice distortions and carrier self-trapping, which significantly reduce the carrier mobility and suppress the overall conductive response, thereby giving rise to negative photoconductivity.

Similar Dirac fermion relaxation processes have also been observed in other Dirac semimetals, such as $ZrTe_5$[130,131], $SrMnSb_2$[132], and $PtSe_2$[133], as well as in Kagome semimetals like $Fe_3Sn_2$[134], $GdMn_6Sn_6$[135], and $CsV_3Sb_5$[136-139]. In these systems, photoexcited coherent phonon modes can actively drive topological phase transitions. Moreover, in some magnetic Dirac semimetals, magnetic order acts as a controllable order parameter that switches the system between Dirac and Weyl points, as exemplified by EuAgAs[140] and $MoB_3$[141]. These cases are discussed in the following sections.

## 4.2 Weyl Semimetals

Similar to Dirac semimetals, the ultrafast carrier dynamics in Weyl semimetals are also primarily governed by electron-hole recombination following photoexcitation[105,107]. additional chirality degree of freedom associated with Weyl fermions gives rise to much richer nonequilibrium dynamics. During the intraband relaxation process, electron-phonon scattering is strongly suppressed because of the extremely low density of states near the Weyl nodes[142,143]. As a consequence, both hot-carrier relaxation and selected coherent phonon excitations exhibit unusually long lifetimes. These characteristics make Weyl semimetals an excellent platform for investigating long-lived nonequilibrium electronic states and coherent many-body excitations.

TaAs provides a representative example. The electronic bands around the Weyl nodes exhibit pronounced electron-hole asymmetry, leading to an asymmetric distribution of photoexcited electrons and holes in momentum space and consequently strongly suppressing electron-hole recombination. Ultrafast pump-probe spectroscopy clearly reveals this topology-controlled relaxation mechanism. As shown in Fig. 8(a), the relaxation dynamics of photoexcited carriers in TaAs exhibit a pronounced dependence on the probe wavelength. When the probe wavelength is far from the Weyl nodes (525 nm), carriers relax rapidly through electron-hole recombination. In contrast, when the probe photon energy is tuned close to the Weyl nodes (560 nm), phase-space blocking strongly suppresses the recombination channel, and carrier relaxation becomes dominated by the much slower electron-phonon scattering process, yielding relaxation times of approximately 190-250 ps[143,144]. A similar wavelength-dependent response has also been observed in NbP single crystals, arising from the combined effects of the wavelength-dependent absorption coefficient and the low density of states around the Weyl nodes[145]. These observations suggest that Weyl semimetals may support population inversion analogous to that found in Dirac semimetals, but with substantially longer lifetimes. Such long-lived nonequilibrium carrier populations are expected to provide favorable conditions for topology-enhanced nonlinear optical phenomena, including high-harmonic generation and terahertz emission, although further experimental verification is still required.

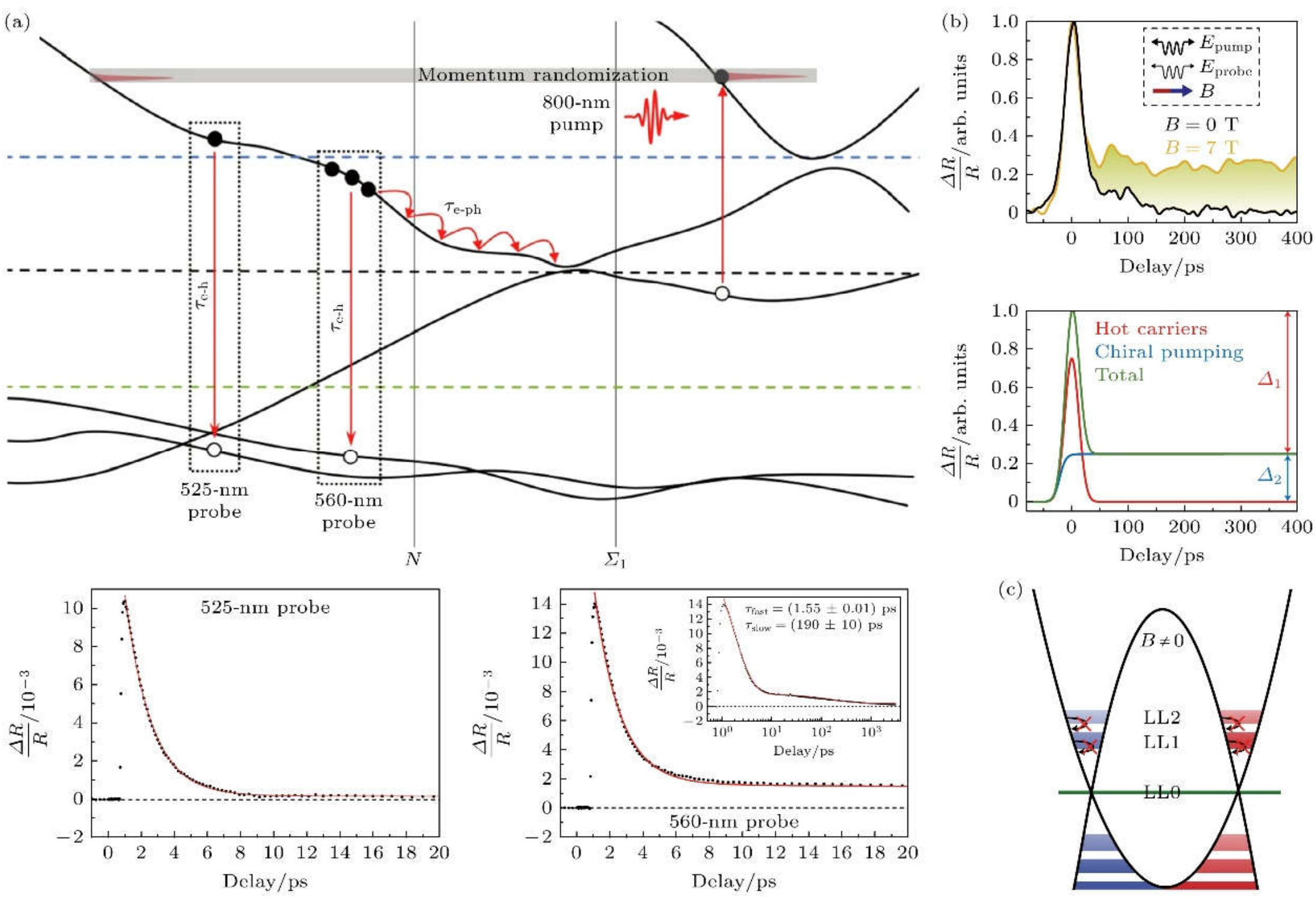

**Fig. 8. Ultrafast dynamics and chiral pumping in the Weyl semimetal TaAs**: (a) Transient reflectivity measured using probe wavelengths away from (525 nm) and close to (560 nm) the Weyl nodes, illustrating the wavelength dependence of carrier relaxation[143]; (b) time evolution of the reflected terahertz probe signal without an external magnetic field and with the magnetic field applied parallel to the terahertz electric field, together with theoretical simulations[147]; (c) schematic illustration of the strong suppression of hot-carrier-phonon scattering by phase-space restrictions in the lowest Landau levels[147].

In three-dimensional Weyl systems, circularly polarized light can break time-reversal symmetry and generate a net imbalance between Weyl nodes of opposite chirality, commonly referred to as the chiral pumping effect[146]. This effect becomes particularly prominent when an external magnetic field is applied parallel to the terahertz electric field. Jadidi *et al*.[147] employed ultrafast optical-pump terahertz-probe spectroscopy using 3.4 THz (88.6 μm, 14 meV) radiation to investigate TaAs. They observed a distinct signature of chiral charge pumping when the magnetic field $B$ was aligned parallel to the terahertz electric field $E$. As shown in Fig. 8(b), in the absence of a magnetic field, photoexcited carriers relax predominantly through electron-acoustic-phonon scattering with a characteristic lifetime of approximately 55 ps. When a magnetic field is applied, however, carrier recombination is strongly suppressed because the separation of Weyl nodes in momentum space, together with the associated chiral selection rules, significantly reduces the available scattering phase space, as illustrated in Fig. 8(c). Consequently, the carrier lifetime increases dramatically to the nanosecond regime or even longer[147].

Type-II Weyl semimetals, including $WTe_2$[148-152], $MoTe_2$[152-154], $TaIrTe_4$[155,156], and magnetic Weyl semimetals such as $Mn_3Sn$[157], CeAlSi[158], and $Co_3Sn_2S_2$[1,159,160], possess strongly tilted Weyl cones, resulting in the coexistence of electron and hole pockets near the Fermi level. This unique electronic structure considerably enhances the topological response of the system and exhibits pronounced sensitivity to the orbital angular momentum of light. Consequently, these materials have attracted considerable attention as promising candidates for high-speed photodetectors and spin-optoelectronic devices.

In $Co_3Sn_2S_2$, time- and spin-resolved ultrafast spectroscopy has provided evidence that the annihilation of Weyl nodes is intimately coupled to the evolution of magnetic order. Furthermore, carriers near the spin-polarized energy gap retain the spin-momentum-locking characteristics of Weyl fermions and exhibit a topology-induced phonon bottleneck effect[160]. Ultrafast MOKE spectroscopy has further revealed strong dynamical coupling between magnetic order and topological electronic states. As shown in Figs. 9(a)–9(c), three characteristic dynamical processes are observed in the ferromagnetic phase ($T < T_c \approx 170$ K)[1]: (i) ultrafast demagnetization on a timescale of approximately 300 fs, dominated by spin-flip scattering; (ii) a slower demagnetization process with a characteristic time of approximately 6 ps, associated with the semimetallic electronic structure; and (iii) an unusual ultrafast magnetization enhancement persisting for several hundred picoseconds. As the temperature increases, a pair of Weyl nodes gradually approach each other and annihilate near 110 K, resulting in the opening of an expanding energy gap[160]. At nearly the same temperature, the ultrafast magnetization-enhancement component disappears, whereas the slow demagnetization channel becomes dominant[1]. Once the temperature exceeds the Curie temperature, the ferromagnetic order is completely lost, and the ultrafast magneto-optical Kerr response correspondingly vanishes[1]. This temperature dependence clearly indicates that the photoinduced magnetization enhancement originates from the spin-filtering effect associated with the tilted Weyl band structure. As illustrated in Fig. 9(d), during the relaxation of photoexcited electrons, the chirality-selective transport through the tilted Weyl cones enables spin-up electrons to propagate more efficiently than spin-down electrons. Consequently, a transient spin-polarized electron pocket is formed, leading to an increase in the density of spin-up states near the Fermi level and, therefore, a transient enhancement of the net magnetization. This mechanism provides a fundamentally new route for realizing optically controlled magnetization without relying on conventional magnetic-field manipulation.

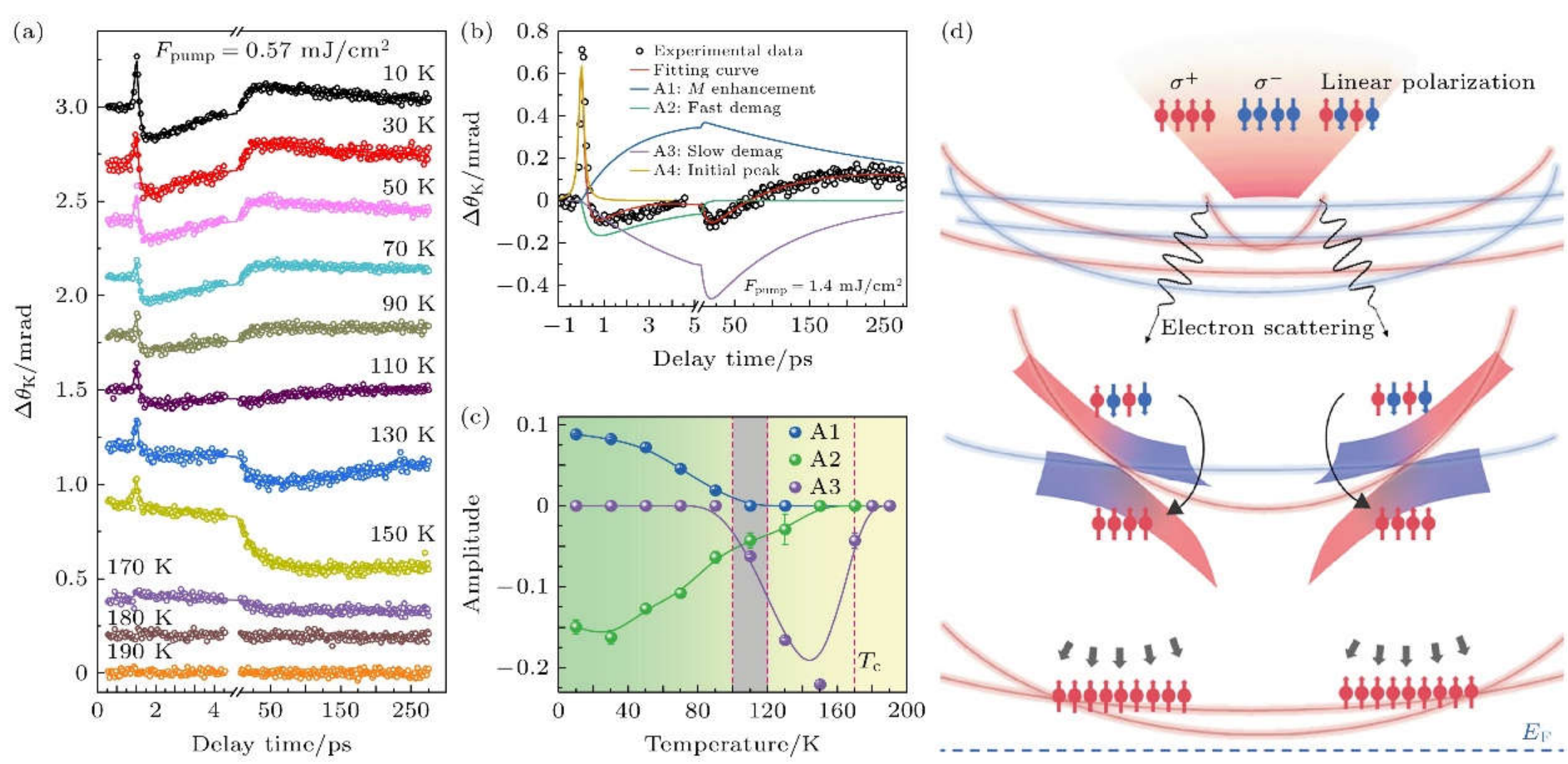

**Fig. 9. Ultrafast magnetization enhancement in $Co_3Sn_2S_2$**[1]: (a) Time evolution of the MOKE angle measured at different temperatures; (b) multi-exponential fitting of the transient Kerr signals; (c) temperature dependence of the fitted relaxation amplitudes; (d) schematic illustration of the spin-filtering effect arising from the tilted Weyl-band structure.

Furthermore, in CeAlSi, the material exhibits a significant nonlinear optical diode effect due to the linear dispersion near the Weyl points[158]. When ultrafast laser pulses propagate in opposite directions, the generated second-harmonic intensities differ by a factor of six. This nonreciprocal response is modulated by the magnetization direction. These findings indicate that Weyl semimetals serve not only as efficient materials for photoelectric conversion but also as foundational platforms for realizing topological nonreciprocal optical devices.

Unlike Dirac and Weyl semimetals, whose conduction and valence bands intersect at isolated zero-dimensional points in momentum space, nodal-line semimetals exhibit continuous band crossings along closed one-dimensional loops protected by crystal symmetry. Representative materials include ZrSiS[161,162], ZrSiTe[163], ZrSiSe[164], LaBi[165], and $PtSn_4$[166]. Ultrafast spectroscopic measurements reveal that the relaxation dynamics of photoexcited carriers in these systems share many similarities with those of Dirac fermions. More importantly, strong electron-phonon coupling together with coherent phonon excitation can drive light-induced Lifshitz transitions, providing direct evidence for ultrafast manipulation of the electronic topology[163]. These results collectively demonstrate that ultrafast laser excitation can serve as an effective nonthermal control parameter for manipulating topological electronic structures and exploring nonequilibrium quantum phenomena.

The studies reviewed above highlight that, although Dirac and Weyl semimetals share similar topological band structures, their nonequilibrium ultrafast dynamics differ substantially. In both classes of materials, photoexcited carriers eventually relax back to the ground state through electron-hole recombination. However, the additional chirality degree of freedom and tunable magnetic order in Weyl semimetals give rise to much richer nonequilibrium phenomena, including chiral pumping, long-lived carrier populations, ultrafast magnetization enhancement, and nonreciprocal nonlinear optical responses. These unique characteristics not only deepen our understanding of the nonequilibrium physics of topological quantum materials but also establish an important foundation for developing ultrafast photodetectors, chiral light sources, and low-power spintronic devices based on topological semimetals.

# 5 Light-Induced Topological Phase Transitions

Ultrafast spectroscopic techniques not only probe nonequilibrium dynamics in topological materials on the intrinsic timescales of interacting charge, spin, orbital, and lattice degrees of freedom, but also actively manipulate topological band structures using intense femtosecond optical pulses. This capability enables the ultrafast creation, manipulation, and erasure of transient quantum states. For example, light-induced symmetry breaking can split a symmetry-protected Dirac node into a pair of Weyl nodes or open a topologically nontrivial gap at the Dirac point. These processes enable transient switching among distinct topological phases, including transitions from Dirac semimetals to Weyl semimetals, topological insulators, and even axion-insulator states. At present, photoinduced topological phase transitions can be broadly classified into three principal categories: (i) electronically driven mechanisms, (ii) lattice-driven mechanisms, and (iii) mechanisms governed by magnetic order. These three pathways correspond to direct reconstruction of the electronic distribution by the optical field, coherent lattice distortions induced by selective phonon excitation, and ultrafast modulation of magnetic order parameters, respectively.

The electric field of an ultrashort laser pulse can directly couple to electronic and spin degrees of freedom, giving rise to Floquet band engineering and nonthermal electronic distributions. Alternatively, resonant excitation of selected phonon modes can induce directional lattice distortions that modify the crystal symmetry and consequently the band topology. In magnetic topological materials, photoexcitation can further manipulate the orientation or magnitude of magnetic order parameters, thereby altering the stability and topology of symmetry-protected electronic states. In the following sections, we review the underlying physical mechanisms and representative experimental progress associated with these three control strategies.

## 5.1 Electronic Mechanisms

Under femtosecond optical excitation, electrons are promoted from the valence band to the conduction band, resulting in a nonequilibrium redistribution of electronic occupation in the vicinity of the Fermi level. Such transient electronic redistribution not only modifies carrier relaxation dynamics but also feeds back into the electronic band structure through many-body interactions, thereby enabling ultrafast control of topological electronic states. One of the most representative approaches is Floquet engineering, in which a periodically driven optical field extends the static Hamiltonian into the Floquet Hilbert space, producing photon-dressed Floquet-Bloch bands. The interaction between electrons and the periodic driving field renormalizes the electronic band structure and modifies the symmetry and topology of the system, thereby enabling reversible switching between distinct topological phases.

Taking the topological insulator $Bi_2Se_3$ as an example, intense mid-infrared radiation with photon energies below the bulk band gap couples directly to the topological surface states, giving rise to Floquet-Bloch bands[167,168]. Theoretical and experimental studies indicate that the crossing behavior of photon-dressed surface bands at specific momentum points depends heavily on the polarization state of the pump light. Specifically, circularly polarized light breaks time-reversal symmetry and induces an additional gap at the Dirac point[167,168].

Beyond Floquet engineering, ultrafast optical excitation can also reshape the adiabatic potential-energy surface of a material, thereby driving coupled structural and topological evolution. For example, in the topological crystalline insulator SnSe, photoexcitation can instantaneously restore the four-fold rotational symmetry ($C_4$) that was disrupted by thermal fluctuations. During the subsequent electron-hole recombination process, the system can relax barrier-free into a topological crystalline-insulator phase protected by the restored crystal symmetry[169]. This concept of photoinduced symmetry restoration substantially broadens the conventional picture of nonequilibrium phase transitions, demonstrating that ultrafast light fields can both break and reconstruct crystal symmetry depending on the excitation pathway.

In topological semimetals, circularly polarized light can dynamically break either time-reversal or inversion symmetry, thereby splitting a fourfold-degenerate Dirac node into two Weyl nodes with opposite chirality[146,170-172]. The number, momentum-space positions, and topological charges of the generated Weyl nodes can be continuously tuned through the polarization, intensity, and frequency of the driving field. For example, in the three-dimensional Dirac semimetal $Na_3Bi$, circularly polarized light generates an effective optically induced magnetic field, driving reversible switching among Dirac-semimetal, Weyl-semimetal, and transient topological-insulator phases, thereby producing a rich light-controlled topological phase diagram[171]. Furthermore, first-principles calculations predict that photoexcitation can induce substantial band renormalization in TaAs, leading to femtosecond shifts in the momentum-space positions of the Weyl nodes[173].

Another important mechanism arises from the intrinsic nonlinear optical response of Weyl semimetals. Owing to their nontrivial Berry curvature and opposite chiralities, Weyl semimetals exhibit chirality-selective second-order nonlinear photocurrents under ultrafast optical excitation. The resulting shift current produces an asymmetric electronic distribution that transiently breaks both time-reversal and inversion symmetries, thereby inducing a purely electronic, nonthermal topological phase transition. In TaAs, femtosecond laser excitation generates ultrafast shift currents that modify the symmetry of the electronic system on the picosecond timescale without any accompanying structural phase transition, demonstrating that the observed topological transition originates entirely from electronic processes[174-176].

A closely related example is provided by $PtTe_2$ thin films containing a controlled gradient of Te vacancies. The defect-induced breaking of inversion symmetry substantially enhances the Berry-curvature-dipole contribution to circular photogalvanic currents and gives rise to helicity-dependent terahertz emission[177]. These findings demonstrate that ultrafast optical fields can manipulate topological electronic responses entirely through electronic degrees of freedom, providing new opportunities for realizing zero-field spin polarization, ultrafast topological photodetection, and nonreciprocal optoelectronic functionalities.

## 5.2 Lattice Mechanisms

In addition to direct electronic excitation, ultrafast optical pulses can resonantly drive specific coherent phonon modes, thereby inducing transient lattice distortions that modify crystal symmetry, electronic band structures, and ultimately the topological phase of the material. Because topological electronic states are often highly sensitive to atomic positions and crystal symmetry, coherent lattice control has emerged as one of the most effective routes for realizing nonthermal topological phase transitions, as illustrated schematically in Fig. 10(a).

In the nodal-line semimetal ZrSiS, ultrafast optical excitation can selectively drive the $A_{1g}^{2}$ coherent phonon mode. The phase of the coherent lattice vibration can be tuned from 0 to π, resulting in sub-picosecond shifts of the Dirac bands and eventually driving a Lifshitz transition[162]. Similarly, in the Dirac semimetal $SrMnSb_2$, first-principles calculations predict that periodic atomic displacements associated with the $A_{1g}$ phonon mode periodically open and close the gap at the Dirac point, demonstrating that coherent lattice vibrations can dynamically regulate the topological band structure[132]. In the Weyl semimetal $(TaSe_4)_2I$, femtosecond laser excitation induces nonthermal melting of both polarons and the charge-density-wave state, thereby continuously driving the system from a polaronic state to the CDW (axion) phase and finally to a hidden Weyl phase[156].

The layered Dirac material $ZrTe_5$ provides another representative example of lattice-driven topological switching. Photoexcitation with 800 nm femtosecond pulses induces transient displacement of Te atoms along the crystallographic *c*-axis, thereby driving a transient transition from a strong topological insulator to a Dirac semimetal[178]. More remarkably, near-single-cycle terahertz pump pulses resonantly excite the Raman-active $A_{1g}$ phonon mode, giving rise to a long-lived metastable state once the excitation exceeds a critical threshold. Theoretical simulations of of the corresponding atomic trajectories demonstrate that coherent excitation of the $A_{1g}$ mode continuously drives $ZrTe_5$ through the sequence of a strong topological insulator, a Dirac semimetal, and finally a weak topological insulator, as illustrated in Fig. 10(b)[130]. Furthermore, optical-pump terahertz-probe spectroscopy has revealed that coherent excitation of the same phonon mode can transiently transform $ZrTe_5$ from a Dirac semimetal into a Weyl semimetal[131].

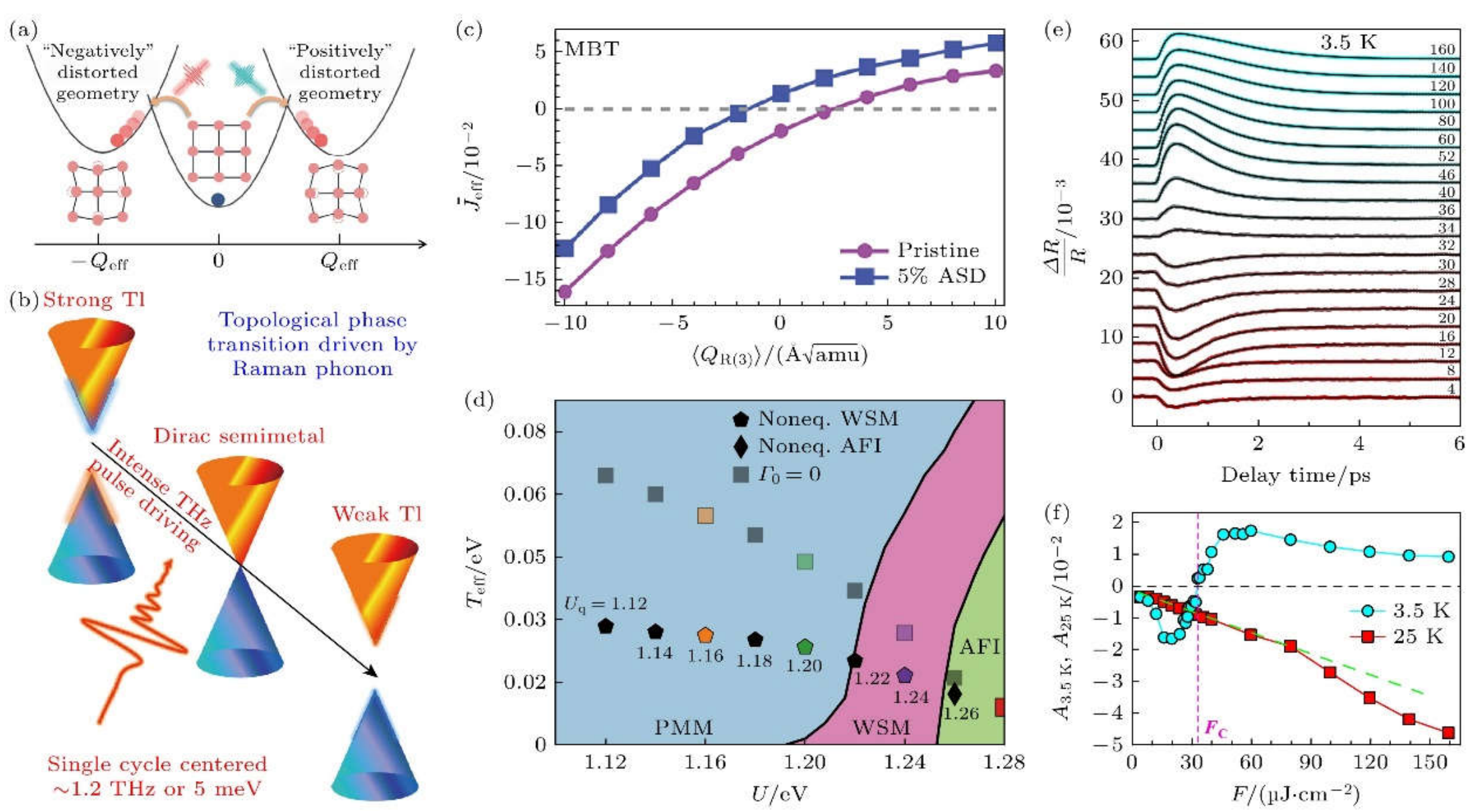

**Fig. 10. Light-induced lattice- and magnetic-order-driven topological phase transitions**. (a) Schematic illustration of coherent control of lattice geometry through photoinduced coherent phonon displacement[162]. (b) Schematic illustration of terahertz-driven coherent phonon excitation inducing topological state switching in $ZrTe_5$[130]. (c) Calculated effective average interlayer exchange interaction in $MnBi_2Te_4$ as a function of the average displacement of the breathing phonon mode $\langle Q_{R(3)} \rangle$, where ASD denotes anti-site defects[92]. (d) Equilibrium phase diagram as a function of temperature and the Hubbard interaction $U$[184]. (e), (f) Pump-fluence-dependent transient reflectivity of EuAgAs measured in the antiferromagnetic phase[140].

Another important lattice-control mechanism originates from the direct modification of the crystal free-energy landscape by the electric field of an ultrafast laser pulse. Following photoexcitation, electrons are promoted into interlayer antibonding orbitals, thereby weakening the interlayer chemical bonding and altering the interlayer coupling strength. The resulting change in the lattice potential drives collective interlayer shear or breathing motion, allowing the crystal to relax toward a new nonequilibrium equilibrium configuration and thereby inducing a light-driven topological phase transition.

For example, this effect has been observed in materials such as TaAs[173,179], $NbIrTe_4$[180], $WTe_2$[148,150,181,182], and $MoTe_2$[3,153,154,183]. Specifically, terahertz pulse-induced interlayer shear strain can transform $WTe_2$ and $MoTe_2$ from the $T_d$ phase Weyl semimetal state to the $1T'$ phase trivial semimetal. At specific photon energies, $WTe_2$ can also regulate Weyl points through orbital-selective electron excitation[150].

## 5.3 Mechanisms Related to Magnetic Order

In magnetic topological materials, the symmetries that protect topological states are jointly determined by the crystal lattice and magnetic order. Therefore, both light-induced lattice distortions and transient magnetic moments can drive topological phase transitions. Ultrafast laser pulses enable precise manipulation of these symmetries, allowing transitions from antiferromagnetic to ferromagnetic states or dynamic switching between different topological phases, for example, from a topological insulator to a Weyl semimetal.

In the two-dimensional antiferromagnetic topological insulators $MnBi_2Te_4$ and $MnSb_2Te_4$, theoretical calculations indicate that excitation of the shear and breathing Raman modes can reverse the sign of the interlayer exchange interaction (Fig. 10(c)). This reversal drives an antiferromagnetic-to-ferromagnetic transition, suggesting the possibility of ultrafast optical control of topological phase transitions[92]. However, current ultrafast experiments on $MnBi_2Te_4$ thin layers are conducted under external magnetic fields. Ultrafast MOKE effect and transient reflectivity measurements have proven effective for tracking magnetic-state transitions, and coherent magnon modes associated with the canted antiferromagnetic phase have been identified[2]. transient reflectivity measurements have proven effective for tracking magnetic-state transitions, and coherent magnon modes associated with the canted antiferromagnetic phase have been identified.

In antiferromagnetic topological materials, the screening effect of photoexcited electrons reduces the magnetic moment, thereby modifying the topological states protected by the magnetic space-group symmetry. As a result, a magnetic Dirac semimetal may evolve into a Weyl semimetal or even a topologically trivial state. In addition, ultrafast laser excitation can dynamically modulate the effective electron-electron interaction by renormalizing the Hubbard $U$, thereby facilitating light-induced topological phase transitions.

As shown in Fig. 10(d), theoretical calculations suggest that laser- or quench-induced excitation dynamically suppresses magnetic order, driving a phase transition from an antiferromagnetic insulator to a Weyl semimetal in pyrochlore iridates[184]. In monolayer $MoB_3$, an ultrafast demagnetization process on the timescale of several hundred femtoseconds can trigger a transition from a Dirac nodal-line semimetal to a metallic state[141]. This ultrafast demagnetization mechanism not only demonstrates the transient control of magnetic order by optical excitation but also provides new insights into the design of ultrafast spintronic devices.

Beyond ultrafast demagnetization, ultrafast rotation of the Néel vector in antiferromagnetic Dirac semimetals can also regulate the gap at the Dirac point, thereby driving a topological phase transition[140,185]. Liu *et al.*[140] identified a possible sub-picosecond photoinduced topological phase transition in EuAgAs, which is closely related to the magnetic structure of the material. The transient reflectivity changes dramatically at the antiferromagnetic transition temperature, indicating that antiferromagnetic order strongly influences the relaxation dynamics of photoexcited carriers, as shown in Figs. 10(e) and 10(f). In the low-temperature antiferromagnetic phase, the ultrafast response exhibits a pronounced pump-fluence dependence. At sufficiently high pump fluence, EuAgAs evolves from an antiferromagnetic Dirac semimetal toward a possible transient Weyl-semimetal state[140].

# 6 Conclusion and Outlook

Ultrafast spectroscopy has become an indispensable tool for investigating the nonequilibrium dynamics of charge, lattice, and spin degrees of freedom in topological materials, owing to its femtosecond-to-nanosecond temporal resolution and broad wavelength tunability. Substantial progress has been achieved in the study of topological insulators, topological semimetals, and magnetic topological materials. These studies have provided deep insights into key phenomena, including surface-bulk coupling, spin-polarization dynamics, coherent phonon excitations, and coherent magnon dynamics. Moreover, three principal mechanisms underlying light-induced topological phase transitions have been established: electronically driven mechanisms based on nonequilibrium carrier redistribution, lattice-driven mechanisms associated with coherent structural distortions, and magnetic-order-driven mechanisms arising from ultrafast manipulation of magnetic order parameters. These advances have significantly improved our understanding of the nonequilibrium behavior of topological quantum states and have laid the foundation for optically controlled topological phase transitions and their potential device applications.

As ultrafast spectroscopy continues to expand its role in the investigation of topological quantum materials, both new opportunities and significant challenges are emerging. Current experimental techniques are still limited by the accessible momentum resolution, pump-photon-energy range, and excitation-fluence window. To further advance this field, future research should focus on the following directions.

(1) Multidimensional ultrafast spectroscopic characterization. Future studies

should integrate time-resolved angle-resolved photoemission spectroscopy (Tr-ARPES), ultrafast X-ray scattering, terahertz time-domain spectroscopy (THz-TDS), and in situ high-pressure ultrafast spectroscopy[186,187] to simultaneously track the coupled dynamics of electronic, lattice, spin, and orbital degrees of freedom, thereby establishing a comprehensive picture of nonequilibrium topological states.

(2) Nonlinear dynamics under strong-field and low-temperature conditions. Future work should explore extreme nonequilibrium topological phenomena, including Floquet-engineered topological states, high-harmonic generation, chiral pumping, nonlinear Hall effects, and other strong-field topological responses, in order to deepen our understanding of ultrafast topological physics under extreme conditions.

(3) Rational design of controllable topological material systems. Material engineering through chemical doping, strain engineering, dimensionality reduction, layer-number control, and heterostructure design will enable programmable manipulation of topological band structures, quantum phases, and photoinduced phase-transition pathways, providing versatile platforms for ultrafast topological devices.

(4) Deeper integration of theoretical simulations and ultrafast experiments. Further development of nonequilibrium many-body theories, Floquet theory, real-time first-principles calculations, and time-dependent density-functional theory will be essential for understanding, predicting, and quantitatively describing ultrafast light-matter interactions and photoinduced topological phase transitions in complex quantum materials.

Continued progress along these directions is expected to enable precise, reversible, and ultrafast manipulation of topological quantum states. The combination of advanced ultrafast spectroscopic techniques with high-quality quantum materials and predictive theoretical frameworks will not only deepen our understanding of nonequilibrium topological phenomena but also accelerate the development of high-speed, low-power topological electronic devices, spintronic technologies, and quantum-information platforms.

Overall, ultrafast spectroscopy is evolving from a technique primarily used to probe nonequilibrium dynamics into a powerful approach for actively controlling topological quantum states. With the continuous development of advanced light sources, multidimensional spectroscopic techniques, and theoretical methods, the field is expected to play an increasingly important role in uncovering novel quantum phenomena and enabling practical applications of topological quantum materials.